\documentclass[12pt]{article}
\usepackage{pst-plot,url,epsf,axodraw}
\newcommand{\beq}{\begin{equation}}
\newcommand{\eeq}{\end{equation}}

\newcommand{\bea}{\begin{eqnarray}}
\newcommand{\eea}{\end{eqnarray}}
\newcommand{\nobody}{\rule{0ex}{1ex}}

\newcommand{\epm}{e^+e^-}
\newcommand{\gv}{\mbox{GeV}}

\newcommand{\ra}{\rightarrow}
\newcommand{\ga}{\gamma}

\newcommand{\eecsmn}{e^+ e^- \ra c \bar{s} \mu^- \bar{\nu}_{\mu}}

\newcommand{\eetbtn}{e^+ e^- \ra t \bar{b} \tau^- \bar{\nu}_{\tau}}
\newcommand{\eecstng}{e^+ e^- \ra c \bar{s} \tau^- \bar{\nu}_{\tau}\gamma}
\begin{document}
\thispagestyle{empty}
\voffset -2cm
\begin{flushright}{\normalsize \rm
DESY 01-143 [hep-ph/0109290]~~\\
September 2001 (rev. Feb 2002)\\
\vspace*{0.5cm}}
\end{flushright}
\begin{center}
{\LARGE\bf Results for all reactions \boldmath{$e^+ e^- \ra 4{\rm f},
4{\rm f}\gamma$} with
           nonzero fermion masses\footnote{Work supported
           in part by the Polish State Committee for Scientific Research
           (KBN) under contract No. 2~P03B~004~18 and by European Commission's
           5-th Framework contract  HPRN-CT-2000-00149.}}\\
\vspace*{2cm}
Fred Jegerlehner$^{\rm a}$ and
Karol Ko\l odziej$^{\rm b}$\vspace{0.5cm}\\
$\nobody^{\rm a}${\small\it
Deutsches Elektronen-Synchrotron DESY, Platanenallee 6, D-15738 Zeuthen,
Germany}\\
$\nobody^{\rm b}${\small\it
Institute of Physics, University of Silesia, ul. Uniwersytecka 4,
PL-40007 Katowice, Poland}
\vspace*{3.5cm}\\
{\bf Abstract}\\
\end{center}
We accomplish our efforts to obtain predictions for all four--fermion
final states of $e^+ e^-$--annihilation and the corresponding
bremsstrahlung reactions which are possible in the framework of the
Standard Model. For this purpose we have developed a program {\tt
ee4f$\gamma$}. Our predictions are valid for fermions of arbitrary
masses and we can obtain results for total cross sections without any
collinear cut. Keeping exact fermion masses is of course required for
top quark production. We give a detailed phenomenological analysis of
fermion mass effects and real photon radiation for all channels of
four--fermion production at LEP-II and next linear collider energies.
\vfill

\\

\vfill
\newpage
\section{Introduction}
Among the most attractive options of facilities at the high energy
frontier of elementary particle physics are high luminosity $\epm$
linear colliders like
TESLA~\cite{Richard:2001rf,Aguilar-Saavedra:2001rg}, the
NLC~\cite{Abe:2001wn} or the JLC~\cite{Abe:2001gc}. However, going to
higher energies and higher luminosity becomes a real challenge for
working out Standard Model (SM) predictions of the adequate precision
because of the dramatically increasing complexity of perturbative
calculations.  Here we consider all four fermion production processes
in electron positron annihilation together with real photon emission
at the tree level.  These processes with 6 and 7 external particles at
the tree level are described by from 10 up to 1008 Feynman diagrams in
a given channel, neglecting the Higgs boson coupling to light fermion flavors,
and the physical cross section is the result of a very
obscure quantum mechanical interference between all these
diagrams. Interesting are of course those cases where the result is
dominated by a few diagrams like in $W$--pair production where we have
three relevant ``signal diagrams''. However, also in these cases which
allow a relatively simple physical interpretation, many other diagrams
may play a role as a background contribution which affects the precise
interpretation of the signal process. The latter give us information
about the gauge boson parameters and the triple or quadruple gauge
couplings. Similarly, the properties of the Higgs boson will be fixed
by its contribution as unstable intermediate state. The most important
physics cases have been reviewed in~\cite{Aguilar-Saavedra:2001rg},
for example. A number of dominant $\epm \ra 4{\rm f}$ channels have
been explored experimentally at LEP-II
(1996-2000)~\cite{unknown:2001xv} and the measurements confirmed SM
predictions at the level of the most relevant $O(\alpha)$ corrections.

In the approximation of massless fermions all possible four fermion
channels $\epm \ra 4{\rm f}$ have been investigated
in~\cite{Berends:1995xn} ({\tt EXCALIBUR}) and those for $\epm \ra
4{\rm f}\gamma$ in~\cite{Denner:1999gp} ({\tt RacoonWW}). Here we
extend these investigations to a calculation with nonzero fermion
masses.  Keeping nonzero fermion masses will be important in cases
where predictions at the 1\% accuracy level are
required~\cite{Jegerlehner:2000wu,Jegerlehner:2001hc}. Finite masses
also provide a natural regularization of distributions which become
singular in the massless limit. Massive calculations thus provide
reliable benchmarks for massless calculations with cuts. The latter
are much simpler and hence much faster than calculations with massive
codes. The hard bremsstrahlung processes are of interest in their own
right and may be used to investigate anomalous $WW\gamma$ and
$WW\gamma\gamma$ couplings, for example.

Our calculation is considered to be a building block (the soft plus
hard bremsstrahlung part) for a complete $O(\alpha)$ calculation of
the processes $\epm \ra 4{\rm f}$. Such calculations have been
attempted in~\cite{Vicini:1998iy} (see also
\cite{Aeppli:1991zp}). This would also extend existing calculations of
$W$--pair production in the double pole approximation
\cite{Denner:2000bj} ({\tt RacoonWW}) and
\cite{Placzek:2001ng} ({\tt KORALW/YFSWW}) (see also
\cite{Fleischer:1995sq} ({\tt EEWW})) which
incorporate the one-loop corrections for production of on--shell
$W$--pairs~\cite{rc} and their subsequent decay into
fermion--pairs~\cite{decay}.

There already exist a number of codes which allow to calculate exact
matrix elements for $\epm \ra 4{\rm f},\;4{\rm f}\gamma$ for massive
fermions. Some of the program packages available are general purpose
packages which allow for an automatic calculation of tree--level
amplitudes and for their numerical evaluation. Known programs, which
may be utilized for tree-level calculations of the kind we are
interested in are: {\tt GRACE/BASES}~\cite{Tanaka:1991wn}, {\tt
MADGRAPH/HELAS}~\cite{Stelzer:1994ta}, {\tt
CompHEP}~\cite{Boos:1994xb} (squared matrix element technique), {\tt
WPHACT}~\cite{Accomando:1997es}, {\tt
NEXT\-CALI\-BUR}~\cite{Berends:2000gj} (initial state radiation
photons generated via the structure function approach), {\tt
HELAC/PHEGAS}~\cite{Kanaki:2000ey} (recursive Dyson-Schwinger equation
approach), {\tt WRAP}~\cite{Montagna:2001uk} ({\tt
ALPHA}~\cite{Caravaglios:1995cd} algorithm) and {\tt
O'Mega/WHIZARD}~\cite{Moretti:2001zz}.  Most of the codes work on the
basis of helicity amplitudes and some use the structure function approach
to generate the photons. Except for {\tt NEXT\-CALI\-BUR}, which is
specialized to $\epm \ra 4{\rm f}$, all other programs allow to
generate and evaluate amplitudes for other type of processes. For
more details and comparisons we refer to the {\it Four Fermion Working
Group Report}~\cite{Grunewald:2000ju}.

Many channels have their own specific problems concerning numerical
stability and/or efficiency and need separate consideration and
optimization. We therefore present, in this paper, a different approach
which is optimized for each individual channel.  We present reference
tables for cross section calculations at $\sqrt{s}$ = 200 GeV and 500
GeV for all leptonic, semi--leptonic and hadronic channels. The
present work completes previous investigations of specific channels
presented in~\cite{Jegerlehner:2000wu,Jegerlehner:2001hc}.

Our calculation has to be extended to include the $O(\alpha)$ virtual
corrections in future.  Precise knowledge of the various channels is
crucial for the precise determination of properties of the unstable
gauge and Higgs bosons as well as to reveal possible anomalous
coupling~\cite{Aguilar-Saavedra:2001rg,Denner:2001vr,Montagna:2001ej}
which might exist beyond the SM.

In the following we outline our calculation, present the numerical
results and end with the conclusions.

\section{Calculation}
The matrix elements of the reactions
\beq
\label{born}
\epm \ra 4f
\eeq
and
\beq
\label{brems}
\epm \ra 4f\gamma
\eeq
are calculated by utilizing the helicity amplitude method described
in~\cite{Jegerlehner:2000wu}. As in~\cite{Jegerlehner:2000wu},
the photon propagator is taken in the Feynman gauge while for the
propagators of the massive gauge bosons we use the unitary
gauge. Constant widths of the electroweak gauge bosons, $\Gamma_W, \Gamma_Z$,
Higgs boson, $\Gamma_H$, and the top quark, $\Gamma_t$ are
introduced through the complex mass parameters
\beq
\label{cmass}
M_V^2=m_V^2-im_V\Gamma_V,  \quad V=W, Z,  \qquad M_H^2=m_H^2-im_H\Gamma_H,  
\qquad M_t=m_t-i\Gamma_t/2,
\eeq
in the corresponding propagators
\beq
\label{props}
\Delta_F^{\mu\nu}(q)=\frac{-g^{\mu\nu}+q^{\mu}q^{\nu}/M_V^2}{q^2-M_V^2}, \qquad
\Delta_F(q)=\frac{1}{q^2-M_H^2}, \qquad 
S_F(q)=\frac{/\!\!\!q+M_t}{q^2-M_t^2},
\eeq
both in the $s$- and $t$-channel Feynman diagrams.
The electroweak mixing parameter is kept real
\bea
\label{rsw2}
\sin^2\theta_W=1-m_W^2/m_Z^2.
\eea
This kind of parametrization is usually referred to as the fixed-width
scheme (FWS). Our results presented in the next section have been
obtained in the FWS. In our program {\tt ee4f$\gamma$}, it is also possible
to define $\sin^2\theta_W$ in terms of the complex masses
of~(\ref{cmass}) as
\bea
\label{csw2}
\sin^2\theta_W=1-M_W^2/M_Z^2,
\eea
which is usually called the complex-mass scheme (CMS). The CMS has the
advantage that it satisfies the $SU(2) \times U(1)$ Ward identities at
tree level~\cite{Denner:1999gp}.  However, the fact that~(\ref{csw2})
makes some of the SM couplings complex quantities may become a source
of discomfort. In the FWS on the other hand, all the couplings remain
real. Only the electromagnetic gauge invariance is satisfied exactly,
however, and this only provided that $\Gamma_t$ and the other fermion
widths are vanishing.  It should be stressed at this point, that for
vanishing fermion widths, electromagnetic gauge invariance is
preserved with non-zero fermion masses and with the gauge boson widths
$\Gamma_W$ and $\Gamma_Z$ treated as independent parameters. If a
non-vanishing top quark width is introduced through
substitution~(\ref{cmass}), which is done in order to regularize the
on-mass-shell pole of a top quark propagator at tree level in
reactions (\ref{brems}) containing a single top quark in the final
state, the external electromagnetic gauge invariance gets violated. It
can be restored by redefining the Dirac bi-spinor representing the
external top quark in such a way that it satisfies the Dirac equation
with the complex top quark mass of Eqs.~(\ref{cmass}), which would
obviously require a complex top quark four momentum in the phase space
generation. As the production of a on-mass-shell top quark is a rather
unphysical process, one should not be to much concerned about the
problem. Fortunately, as we shall see in the next section, the
violation of the gauge symmetry does not lead to dramatic effects for
the total cross sections of reactions~(\ref{brems}) containing a top
quark in the final state. A more realistic treatment of the top quark
has to include its decay and thus requires the consideration of
$e^+e^- \to 6f, 6f\gamma$ channels, which is beyond the task of the
present investigation.

The hadronic channels are discussed only at the level of quark parton
production. Quark-mass effects will be estimated by adopting the
so--called current-quark masses in the $\overline{\mathrm{MS}}$ scheme
at a scale $\mu \sim $ 2 GeV. This should allow us to get an idea
about the size of mass effects and eventually allow us to establish
suitable cuts which eliminate the mass sensitivity of quark production
cross sections. In any case, taking into account mass effects,
provides an improvement over calculations in the approximation of
massless quarks. For observables which exhibit a substantial mass
dependence of course one would have to discuss more carefully the
precise physical meaning of quark masses in the given process.

\bigskip
\section{Results}
In this section, we will present numerical results for all the
four--fermion channels of reactions (\ref{born}) and (\ref{brems})
which are possible in the SM.

We define the SM physical parameters in terms of the gauge boson
masses and widths, the top mass and width, and the Fermi coupling
constant. We take the actual values of the parameters
from~\cite{Groom:2000in}:\\[4mm]
\centerline{
$m_W=80.419\; {\rm GeV}, \quad \Gamma_W=2.12\; {\rm GeV}, \qquad
m_Z=91.1882\; {\rm GeV}, \quad \Gamma_Z=2.4952\; {\rm GeV}$,}
\beq
\label{params}
m_t=174.3\; {\rm GeV}, 
\quad G_{\mu}=1.16639 \times 10^{-5}\;{\rm GeV}^{-2}.
\eeq
We assume a Higgs boson mass of $m_H=115$ GeV and calculate the Higgs
boson width with the lowest order formula in the SM. The top quark width
is assumed to be $\Gamma_t=1.5$~GeV.

For the sake of definiteness we also list the other fermion masses that we
use in the calculation~\cite{Groom:2000in}:\\[4mm]
\centerline{
$m_e=0.510998902\; {\rm MeV}, \quad m_{\mu}=105.658357\; {\rm MeV},\quad
m_{\tau}=1777.03\; {\rm MeV}$,}
\beq
m_u=5\; {\rm MeV}, \quad m_d=9\; {\rm MeV}, \quad m_s=150\; {\rm MeV}, \quad
m_c=1.3\; {\rm GeV}, \quad m_b=4.4\; {\rm GeV}.
\label{fmass}
\eeq
We neglect the Cabibbo--Kobayashi--Maskawa (CKM) mixing, i.e., we
assume the CKM matrix to be the unit matrix. However, it is possible to
run the program with nontrivial CKM mixing as well.

The effective fine structure constant (at scale $\sim \;M_W$) is
calculated via
\beq
\label{alphaw}
\alpha_W=\sqrt{2} G_{\mu} m_W^2 \sin^2\theta_W/\pi
\eeq
utilizing the real electroweak mixing parameter $\sin^2\theta_W$
defined by~(\ref{rsw2}). In {\tt ee4f$\gamma$}, it is also possible to 
perform computations
with the complex $\sin^2\theta_W$ of Eq.~(\ref{csw2}) and the complex
$m_W^2$ of Eqs.~(\ref{cmass}), i.e. with the complex $\alpha_W$.
The photon coupling to fermions and gauge bosons is given by
the fine structure constant in the Thomson limit
$\alpha=1/137.0359895$ and the quark--gluon strong interaction
``constant'' by $\alpha_s(M_Z)=0.1185$.

We will apply the following set of ``standard cuts'' which have been proposed
in~\cite{Grunewald:2000ju}:
\beq
\begin{array}[b]{cccc}
\cos\theta (l,\mathrm{beam})\le 0.985, & \theta( \gamma, l)> 5^\circ,
& E_\ga> 1\;\gv,  & m(q,q')> 10\;\gv\,, \\
\cos\theta (\gamma,\mathrm{beam})\le 0.985, & \theta( \gamma, q)> 5^\circ,
& E_l> 5\;\gv,
\end{array}
\label{cancuts}
\eeq
where $l$, $q$, $\gamma$, and ``beam'' denote charged leptons, quarks,
photons, and the beam (electrons or positrons), respectively, and
$\theta(i,j)$ the angles between the particles $i$ and $j$ in the
center of mass system. Furthermore, $m(q,q')$ denotes the
invariant mass of a quark pair $qq'$. Note that we are not applying a
corresponding cut to the invariant mass of the charged lepton pairs.

The errors we will quote in the Tables below have been evaluated as
follows: For each separate channel of the multichannel Monte Carlo
(MC) integration the error is calculated by {\tt
VEGAS}~\cite{Lepage:1978sw}. This is a purely statistical error
equivalent to one standard deviation. We added linearly standard
deviations for all the channels used in an integration and this is
what is our error. This provides a more conservative estimate for the
error than for example adding up partial errors in quadrature.

\begin{table}
{\small Table~1: Cross sections in femto-barns (1 fb=$10^{-15}$ barns)
of $\eecstng$ and $\eetbtn \gamma$ in CMS at $\sqrt{s}=500$ GeV for different
photon energy cuts $\omega$.  $\sigma_s$, corresponding to $E_{\gamma}
\le \omega$, is the cross section in the soft photon limit and
$\sigma_h$, corresponding to $E_{\gamma} > \omega$, is the hard
bremsstrahlung cross section.  A fictitious photon mass
$m_{\gamma}=10^{-6}$ GeV has been introduced in order to regularize
the infrared divergence.  No other cuts except for $\omega$ and
$m_{\gamma}$ are present.}

\begin{center}
\begin{tabular}{|c|c|c|c|c|c|c|}
\hline
\rule{0mm}{7mm} $\omega$ & \multicolumn{3}{c|}{$\eecstng$}
                         & \multicolumn{3}{c|}{$\eetbtn \gamma$} \\[2mm]
\cline{2-7}
\rule{0mm}{7mm} (GeV)
    & $\sigma_s$ (fb) & $\sigma_h$ (fb) & $\sigma_s + \sigma_h$ (fb)
     & $\sigma_s$ (fb) & $\sigma_h$ (fb) & $\sigma_s + \sigma_h$ (fb) \\[2mm]
\hline
\rule{0mm}{7mm}0.1 & 186.37(9) & 250.7(3) & 437.1
                               & 51.85(3)  & 56.88(9) & 108.7 \\[1.5mm]
              0.01 & 108.89(6) & 328.1(3) & 437.0
                               & 32.29(2)  & 76.1(1)  & 108.4 \\[1.5mm]
             0.001 & 31.39(2)  & 405.5(4) & 436.9
                               & 12.729(8) & 95.3(1)  & 108.0 \\[1.5mm]
\hline
\end{tabular}
\end{center}
\end{table}

Except for the check of electromagnetic gauge invariance, discussed in
the previous section, we perform a few other checks. Whenever the
fermion masses play no role we have reproduced the results
of~\cite{Denner:1999gp}. The matrix elements of almost all channels of
the processes (\ref{born}) and (\ref{brems}) under consideration have
been checked against {\tt MADGRAPH}~\cite{Stelzer:1994ta}. The
comparison was not simple for the channels involving a gluon exchange,
since the version of {\tt MADGRAPH} which we are using, generates
either the electroweak or the QCD part, but not both
simultaneously. In addition, for the reaction $\epm \ra
\epm\epm\gamma$ {\tt MADGRAPH} generates only 999 instead of all 1008
Feynman graphs. The phase space generation routines have been
thoroughly checked against each other before they have been combined
into a multichannel phase space generation routine. The total cross
sections of the reactions (\ref{born}) containing a single top quark in
the final state
\beq
\label{singtop}
         \epm \ra t \bar{b} f \bar{f'},
\eeq
where $f = e^-,\mu^-,\tau^-,d,s$ and $f' =
\nu_{e},\nu_{\mu},\nu_{\tau},u,c$, respectively, have been calculated
in an arbitrary linear gauge~\cite{Biernacik:2001mp}. This does not allow 
to estimate the absolute size of gauge violation effects caused by the 
nonzero widths of the unstable fermions. However, as the transition between
two linear gauges, the 't Hooft--Feynman and unitary gauge, which has been
numerically performed by changing the gauge parameter from 1 to $10^{16}$
has caused practically negligible change in the cross sections at center of
mass energies up to 2 TeV, typical for a linear collider, one may expect
that the gauge violation effects are not very dramatic for total
cross sections. 

Moreover, another test of reliability of our results
has been performed for several final states. We have split the cross 
section of the bremsstrahlung process
(\ref{brems}) into a soft photon part $\sigma_s$, which includes the
photons with energies $E_{\gamma} \le \omega $, and a hard photon part
$\sigma_h$, including the contributions from photon with energies
$E_{\gamma} > \omega $, and checked whether the combined
bremsstrahlung cross section $\sigma_\gamma=\sigma_s+\sigma_h$ is
independent of the photon energy cut $\omega$
\cite{Jegerlehner:2000wu}.  In Table~1, we illustrate this
independence, which holds within one standard deviation of the MC
integration, for $\eecstng$ in the CMS.  At the same time, in Table~1, we see a
small dependence on the cut--off parameter $\omega$, which is at the
level of about two standard deviations, for the bremsstrahlung
reaction $\eetbtn \gamma$.  The cut dependence is most probably induced by 
the violation of external electromagnetic gauge invariance caused by the 
nonzero top quark width, which has been introduced to 
$\eetbtn \gamma$ in a somewhat asymmetric way, related to the fact
that the top quark is regarded as on-mass-shell particle at the same
time when the anti-top quark decays. It should be stressed, that the soft 
bremsstrahlung cross sections $\sigma_s$ presented in Table~1 are unphysical, 
as they contain the unphysical photon mass $m_{\gamma}$. They are only 
given in order to show that the leading logarithmic contributions are 
treated properly within the CMS for reactions which do not contain nonzero
fermion width and to illustrate the size of cut off dependence caused
by the nonzero top quark width.
As the cut dependence in $\eetbtn \gamma$ is of the order of 1\% of the
corresponding four fermion Born cross section, one should certainly 
elaborate more on this issue in future, in the context of a more realistic 
six fermion reactions, which would treat the decay of the top quark on 
the same footing as that of the anti-top quark.

\begin{table}
{\small Table~2: Cross sections in fb of $\eecsmn$ at $\sqrt{s}=200$
GeV for different cuts on the photon angle with respect to the quarks
$\theta(\gamma,q)$ or muon $\theta(\gamma,\mu)$ and the remaining cuts
as in~(\ref{cancuts}). Here we use physical parameters
of~\cite{Montagna:2001uk} and parametrize the photon couplings by
$\alpha=1/137.0359895$.}
\begin{center}
\begin{tabular}{|c|c|c|c|}
\hline
\rule{0mm}{7mm} $\theta(\gamma,q)$  & $\theta(\gamma,\mu)$ &
                         \cite{Montagna:2001uk} & Present work \\[2mm]
\hline
\rule{0mm}{7mm} $5^\circ$ & $5^\circ $ & 74.294(29) & 74.267(60) \\
                $1^\circ$ & $1^\circ $ & 93.764(37) & 93.70(7) \\
                $5^\circ$ & $1^\circ $ & 90.157(36) & 90.13(7) \\
              $5^\circ$ & $0.1^\circ $ & 104.777(46) & 104.78(7) \\
                $5^\circ$ & $0^\circ $ & 105.438(45) & 105.48(7) \\
\hline
\end{tabular}
\end{center}
\end{table}

\begin{table}
{\small Table~3: $\epm \ra 4{\rm f}$ cross sections $\sigma$ and
$\epm \ra 4{\rm f}\gamma$ cross sections $\sigma_\gamma$ in fb at
$\sqrt{s}=200$ GeV and $\sqrt{s}=500$ GeV for different four--fermion
final states corresponding to the $W^+W^-$--pair signal. The cuts are
those of~(\ref{cancuts}).}
\begin{center}
\begin{tabular}{|c|c|c|c|c|}
\hline
\rule{0mm}{7mm} Final & \multicolumn{2}{c|}{$\sqrt{s}=200$  GeV}
                         & \multicolumn{2}{c|}{$\sqrt{s}=500$  GeV}\\[2mm]
\cline{2-5}
\rule{0mm}{7mm} state
    & $\sigma$   & $\sigma_{\gamma}$ & $\sigma$ & $\sigma_{\gamma}$ \\[2mm]
\hline
\rule{0mm}{7mm}  $u \bar{d} \mu^- \bar{\nu}_{\mu}$         
    & 630.65(31) & 70.547(83) & 211.11(13) & 23.601(46) \\[1.5mm]
$u \bar{d} \tau^- \bar{\nu}_{\tau}$                        
    & 630.18(31) & 68.321(74) & 210.95(13) & 23.386(44) \\[1.5mm]
$c \bar{s} \mu^- \bar{\nu}_{\mu}$                          
    & 630.40(31) & 69.501(80) & 211.03(13) & 23.285(47) \\[1.5mm]
$c \bar{s} \tau^- \bar{\nu}_{\tau}$                        
    & 629.93(31) & 67.279(72) & 210.87(13) & 23.077(43) \\[1.5mm]
$t \bar{b} \mu^- \bar{\nu}_{\mu}$                          
    &  --        &    --      & 58.88(30)  & 9.467(60)  \\[1.5mm]
$t \bar{b} \tau^- \bar{\nu}_{\tau}$                        
    &  --        &    --      & 58.80(29)  & 9.295(56)  \\[1.5mm]
$c \bar{s} d \bar{u}$                                      
    & 1838.6(1.4) & 172.74(28) & 749.07(50) & 68.34(23) \\[1.5mm]
$t \bar{b} d \bar{u}$                                      
    &  --        &    --      & 177.8(1.9) &  25.55(42) \\[1.5mm]
$t \bar{b} s \bar{c}$                                      
    &  --        &    --      & 177.4(1.9) &  25.37(37) \\[1.5mm]
$\nu_{\tau} \tau^+ \mu^- \bar{\nu}_{\mu}$                  
    &  205.88(15) & 25.784(44) & 60.762(62) & 7.842(23) \\[1.5mm]
$u \bar{d} d \bar{u}$                                      
    & 1921.4(7) & 188.19(46) & 780.66(25) &  74.99(28) \\[1.5mm]
$c \bar{s} s \bar{c}$                                      
    & 1925.7(8) & 184.07(46) & 782.62(28) &  73.46(25) \\[1.5mm]
$t \bar{b} b \bar{t}$                                      
    & -- & -- & 0.85519(56) & 0.073748(78)  \\[1.5mm]
$\nu_{\mu} \mu^+ \mu^- \bar{\nu}_{\mu}$                    
    & 218.91(19) & 28.232(55) & 63.933(70) & 8.475(26) \\[1.5mm]
$\nu_{\tau} \tau^+ \tau^- \bar{\nu}_{\tau}$                
    & 214.94(20) & 26.280(50) & 63.468(74) & 8.299(20) \\[1.5mm]
$\nu_e e^+ e^- \bar{\nu}_e$                                
    & 259.55(31) & 32.012(93) & 195.22(42) & 24.85(14) \\[1.5mm]
\hline
\end{tabular}
\end{center}
\end{table}

We compare our results for the total cross sections of $\eecsmn$ at
$\sqrt{s}=200$ GeV with those of~\cite{Montagna:2001uk} in Table~2.
As in~\cite{Montagna:2001uk}, different cuts on the photon angle with
respect to the quarks $\theta(\gamma,q)$ or muon $\theta(\gamma,\mu)$
are imposed while the remaining cuts are those given
in~(\ref{cancuts}). For the comparison we use the physical
parameters of~\cite{Montagna:2001uk}, i.e.  $G_{\mu}=1.16637 \times
10^{-5}\;{\rm GeV}^{-2}$, $m_Z=91.1867$ GeV, $m_W=80.35$ GeV,
$\sin^2\theta_W=1-m_W^2/m_Z^2$, $\Gamma_Z=2.49471$ GeV,
$\Gamma_W=2.04277$ GeV, $m_{\mu}=0.10565839$ GeV, $m_s=0.15$ GeV,
$m_c=1.55$ GeV. Although it has not been explicitly specified, we
assume that \cite{Montagna:2001uk} is using $\alpha=1/137.0359895$ for
the photon coupling strength. We find that the results agree perfectly
within one standard deviation of the MC integration.

Our results for all channels of reactions $\epm \ra 4{\rm f},\;4{\rm
f}\gamma$ possible in the SM are collected in Tables 3--7. We present
total cross sections at two center of mass energies, $\sqrt{s}=200$
GeV and $\sqrt{s}=500$ GeV, with cuts defined by~(\ref{cancuts}),
except for $e^+e^- \ra e^+e^-e^+e^-$, where we have imposed another
cut on the angle between the final state electrons and/or positrons,
$\theta(e^{\pm},e^{\pm}) > 5^\circ $.

\begin{table}
{\small Table~4: $\epm \ra 4{\rm f},\;4{\rm f}\gamma$ cross sections
in fb at $\sqrt{s}=200$ GeV and $\sqrt{s}=500$ GeV for different
four--fermion final states corresponding to the single--$W$ signal. The
cuts are those of~(\ref{cancuts}).}
\begin{center}
\begin{tabular}{|c|c|c|c|c|}
\hline
\rule{0mm}{7mm} Final & \multicolumn{2}{c|}{$\sqrt{s}=200$  GeV}
                         & \multicolumn{2}{c|}{$\sqrt{s}=500$  GeV}\\[2mm]
\cline{2-5}
\rule{0mm}{7mm} state
    & $\sigma$   & $\sigma_{\gamma}$ & $\sigma$ & $\sigma_{\gamma}$ \\[2mm]
\hline
\rule{0mm}{7mm}  $u \bar{d} e^- \bar{\nu}_e$               
    & 661.68(40) & 72.95(10) & 354.06(27) & 38.876(88) \\[1.5mm]
$c \bar{s} e^- \bar{\nu}_e$                                
    & 661.42(40) & 71.84(10) & 353.91(27) & 38.371(88) \\[1.5mm]
$t \bar{b} e^- \bar{\nu}_{e}$                              
    & -- & -- & 58.54(65)   & 9.43(16) \\[1.5mm]
$\nu_{\mu} \mu^+ e^- \bar{\nu}_{e}$                        
    & 216.29(21) & 27.473(57) & 107.34(14) & 13.677(43) \\[1.5mm]
$\nu_{\tau} \tau^+ e^- \bar{\nu}_{e}$                      
    & 216.13(21) & 26.709(55) & 107.25(13) & 13.471(42) \\[1.5mm]
\hline
\end{tabular}
\end{center}
\end{table}

In Tables~3 and 4, we show the results for the channels corresponding
to the $W^+W^-$--pair and single--$W$ signal. These channels are usually
classified as charged current reactions. The relative magnitude of the
cross sections in both tables reflects the naive counting of the color
degrees of freedom, e.g., the cross sections of purely hadronic
channels are about a factor 3 bigger than the cross sections of
semi-leptonic channels and the latter are a factor 3 bigger than the
cross sections of purely leptonic reactions. This somewhat general
rule is obviously violated in reactions which receive contributions
from the gluon or $t$-channel photon and $Z$ exchange. Except for the
channels containing heavy quarks, $t$ and $b$, for lighter flavors,
the fermion mass effects are not big. However, for individual channels
they are of the order of a few per cent, as it has been already
pointed out in~\cite{Jegerlehner:2001hc}. It is amazing that the mass
effect is inverse for $e^+e^- \ra u \bar{d} d \bar{u}$ and $e^+e^- \ra
c \bar{s} s \bar{c}$. The inversion is not due to the Higgs boson exchange,
but in fact is a consequence of the
cuts~(\ref{cancuts}) which we imposed. The latter reduce the
contribution of the $s$-channel Feynman diagrams to $e^+e^- \ra u
\bar{d} d \bar{u}$ to much larger extent than to $e^+e^- \ra c \bar{s}
s \bar{c}$, because the cuts on the invariant mass of the quark pairs
restrict the phase space much more severely for lighter quarks than
for heavier ones.
Without the cuts, the cross section of $\epm \ra u \bar{d} d \bar{u}$ 
becomes bigger than that of $e^+e^- \ra c \bar{s} s \bar{c}$, as
expected.
We do not show cross sections at $\sqrt{s}=200$ GeV for reactions
containing a $t$-quark in the final states, as they are negligibly small
\cite{Biernacik:2001mp}.

{\small

\begin{table}[h]
{\small Table~5: Cross sections in fb of the purely leptonic
neutral-current channels of (\ref{born}) and (\ref{brems})
at $\sqrt{s}=200$ GeV and $\sqrt{s}=500$ GeV.
The cuts are given by~(\ref{cancuts})
with the exception of $e^+e^- \ra e^+e^-e^+e^-$, where we have imposed
another cut on the angle between the final state electrons and/or positrons,
$\theta(e^{\pm},e^{\pm}) > 5^\circ $.}
\begin{center}
\begin{tabular}{|c|c|c|c|c|}
\hline
\rule{0mm}{7mm} Final & \multicolumn{2}{c|}{$\sqrt{s}=200$  GeV}
                         & \multicolumn{2}{c|}{$\sqrt{s}=500$  GeV}\\[2mm]
\cline{2-5}
\rule{0mm}{6mm} state
    & $\sigma$   & $\sigma_{\gamma}$ & $\sigma$ & $\sigma_{\gamma}$ \\[1.5mm]
\hline
\rule{0mm}{7mm}
$\mu^+ \mu^- \tau^+ \tau^- $                                
    & 10.267(14) & 2.1787(91) & 2.5117(44) & 0.6495(40) \\[1.5mm]
$\mu^+ \mu^- \bar{\nu}_{\tau} \nu_{\tau} $                  
    & 12.729(10) & 1.6998(44) & 3.0336(31) & 0.5695(20) \\[1.5mm]
$\tau^+ \tau^- \bar{\nu}_{\mu} \nu_{\mu} $                  
    & 9.1659(59) & 1.2307(27) & 2.7174(27) & 0.5221(15) \\[1.5mm]
$\bar{\nu}_{\tau} \nu_{\tau}\bar{\nu}_{\mu} \nu_{\mu} $     
    & 10.608(6) & 0.57713(82) & 3.9179(43) & 0.4645(11) \\[1.5mm]
$ \mu^+ \mu^- e^+ e^- $                                     
    & 137.18(90) & 12.93(31) & 43.80(38) & 4.58(12) \\[1.5mm]
$ \tau^+ \tau^- e^+ e^- $                                   
    & 54.49(18) & 7.115(55) & 16.860(61) & 2.685(23) \\[1.5mm]
$\mu^+ \mu^- \bar{\nu}_{e} \nu_{e} $                        
    & 17.780(18) & 2.1467(43) & 24.031(92) & 3.479(20) \\[1.5mm]
$\tau^+ \tau^- \bar{\nu}_{e} \nu_{e} $                      
    & 11.721(10) & 1.4797(24) & 21.389(90) & 3.432(21) \\[1.5mm]
$ \bar{\nu}_{\mu} \nu_{\mu} \bar{\nu}_{e} \nu_{e} $         
    & 11.448(8) & 0.6033(7) &  24.728(47) & 2.349(7)   \\[1.5mm]
$ \bar{\nu}_{\mu} \nu_{\mu} e^+ e^- $                       
    & 23.503(22) & 2.7789(68) & 9.453(14) & 1.3076(57) \\[1.5mm]
$ \mu^+ \mu^- \mu^+ \mu^- $                                 
    & 6.747(13) & 1.4307(84) & 1.5106(36) & 0.3860(30) \\[1.5mm]
$ \tau^+ \tau^- \tau^+ \tau^- $                             
    & 3.7283(30) & 0.7943(21) & 1.0341(11) & 0.2727(8) \\[1.5mm]
$ \bar{\nu}_{\mu} \nu_{\mu} \bar{\nu}_{\mu}\nu_{\mu} $      
    & 5.2660(23) & 0.28562(25) & 1.9577(15) & 0.23149(34) \\[1.5mm]
$ e^+ e^- e^+ e^- $                                         
    & 50.53(10)  & 5.770(58) & 13.927(32) & 2.163(21) \\[1.5mm]
$ \bar{\nu}_e \nu_e \bar{\nu}_e \nu_e $                     
    &  5.9815(27) & 0.30563(24) & 22.482(61) & 2.0856(92) \\[1.5mm]
\hline
\end{tabular}
\end{center}
\end{table}
}

{\small

\begin{table}
{\small Table~6: Cross sections in fb of the semi-leptonic
neutral-current channels of
(\ref{born}) and (\ref{brems}) at $\sqrt{s}=200$ GeV and $\sqrt{s}=500$ GeV.
The cuts are those specified by~(\ref{cancuts}).}
\begin{center}
\begin{tabular}{|c|c|c|c|c|}
\hline
\rule{0mm}{7mm} Final & \multicolumn{2}{c|}{$\sqrt{s}=200$  GeV}
                         & \multicolumn{2}{c|}{$\sqrt{s}=500$  GeV}\\[2mm]
\cline{2-5}
\rule{0mm}{6mm} state
    & $\sigma$   & $\sigma_{\gamma}$ & $\sigma$ & $\sigma_{\gamma}$ \\[1.5mm]
\hline
\rule{0mm}{7mm}  $\bar{u} u \mu^+ \mu^- $                   
    & 27.341(30) & 4.874(13)  &  6.855(10) & 1.5630(62) \\[1.5mm]
$\bar{u} u \tau^+ \tau^- $                                  
    & 19.543(17) & 3.4694(98) &  5.9096(69) & 1.3614(49) \\[1.5mm]
$\bar{u} u \bar{\nu}_{\mu} \nu_{\mu} $                      
    & 22.614(14) & 2.2049(34) & 8.3860(99) & 1.2897(30) \\[1.5mm]
$\bar{c} c \mu^+ \mu^- $                                    
    & 27.287(27) & 4.826(19) & 6.9946(88) & 1.5924(74)  \\[1.5mm]
$\bar{c} c \tau^+ \tau^- $                                  
    & 19.560(16) & 3.4407(92) &  6.0566(68) & 1.3922(48) \\[1.5mm]
$\bar{c} c \bar{\nu}_{\mu} \nu_{\mu} $                      
    & 22.655(13) & 2.1695(44) & 8.6652(83) & 1.3434(39) \\[1.5mm]
$\bar{t} t \mu^+ \mu^- $                                    
    & -- & -- &  0.1832(2)  & 0.02165(5) \\[1.5mm]
$\bar{t} t \tau^+ \tau^- $                                  
    & -- & -- & 0.11991(10) & 0.01554(2) \\[1.5mm]
$\bar{t} t \bar{\nu}_{\mu} \nu_{\mu} $                      
    & -- & -- & 0.07010(2)& 0.004577(4) \\[1.5mm]
$\bar{d} d \mu^+ \mu^- $                                    
    & 29.541(24) & 4.229(13)  & 7.0683(80) & 1.3715(60) \\[1.5mm]
$\bar{d} d \tau^+ \tau^- $                                  
    & 21.256(16) & 3.0660(51) & 6.2854(61) & 1.2494(26) \\[1.5mm]
$\bar{d} d \bar{\nu}_{\mu} \nu_{\mu} $                      
    & 24.634(15) & 1.5960(28) & 9.0094(91) & 1.1230(33) \\[1.5mm]
$\bar{s} s \mu^+ \mu^- $                                    
    & 29.542(25) & 4.240(14)  & 7.0702(73) & 1.3737(55)  \\[1.5mm]
$\bar{s} s \tau^+ \tau^- $                                  
    & 21.257(16) & 3.0666(76) &  6.2907(62) & 1.2489(38) \\[1.5mm]
$\bar{s} s \bar{\nu}_{\mu} \nu_{\mu} $                      
    & 24.637(15) & 1.5975(29) & 9.0125(89) & 1.1301(29) \\[1.5mm]
$\bar{b} b \mu^+ \mu^- $                                    
    & 29.536(23) & 4.150(13) & 8.6899(74) & 1.7290(56) \\[1.5mm]
$\bar{b} b \tau^+ \tau^- $                                  
    & 21.349(16) & 3.0179(76)  & 7.9184(66) & 1.5972(43) \\[1.5mm]
$\bar{b} b \bar{\nu}_{\mu} \nu_{\mu} $                      
    & 25.019(15) &  1.5457(28) & 12.322(9) & 1.5911(34) \\[1.5mm]
$ \bar{u} u e^+ e^- $                                       
    & 77.31(26)  & 10.438(74)  & 28.37(11) & 4.432(33) \\[1.5mm]
$ \bar{c} c e^+ e^- $                                       
    & 77.02(19)  & 10.321(61) & 29.030(91) & 4.532(32) \\[1.5mm]
$ \bar{t} t e^+ e^- $                                       
    & -- & -- & 172.63(30) & 15.657(45) \\[1.5mm]
$ \bar{d} d e^+ e^- $                                       
    & 58.759(82) & 7.308(27) & 23.668(46) & 3.363(21) \\[1.5mm]
$ \bar{s} s e^+ e^- $                                       
    & 58.734(80) & 7.337(26) & 23.702(46) & 3.377(18) \\[1.5mm]
$ \bar{b} b e^+ e^- $                                       
    & 58.082(66) & 7.125(21) & 29.38(11) & 4.152(27) \\[1.5mm]
$ \bar{u} u \bar{\nu}_e \nu_e $                             
    & 25.502(19) & 2.4289(41) & 49.18(22) & 6.300(38) \\[1.5mm]
$ \bar{c} c \bar{\nu}_e \nu_e $                             
    & 25.907(21) & 2.4210(38) & 55.13(26) & 7.180(47) \\[1.5mm]
$ \bar{t} t \bar{\nu}_e \nu_e $                             
    & -- & -- & 0.07400(4) & 0.004810(4) \\[1.5mm]
$ \bar{d} d \bar{\nu}_e \nu_e $                             
    & 26.820(20) & 1.6916(23) & 57.03(27) & 5.731(36) \\[1.5mm]
$ \bar{s} s \bar{\nu}_e \nu_e $                             
    & 26.824(22) & 1.6955(24) & 57.320(28) & 5.777(41) \\[1.5mm]
$ \bar{b} b \bar{\nu}_e \nu_e $                             
    & 31.346(28) & 1.8819(33) & 128.31(65) & 12.59(12) \\[1.5mm]
\hline
\end{tabular}
\end{center}
\end{table}
}

{\small

\begin{table}[t]
{\small Table~7: Cross sections in fb of the purely  hadronic
neutral-current channels of (\ref{born}) and (\ref{brems})
at $\sqrt{s}=200$ GeV and $\sqrt{s}=500$ GeV.
The cuts are those specified by~(\ref{cancuts}).}
\begin{center}
\begin{tabular}{|c|c|c|c|c|}
\hline
\rule{0mm}{7mm} Final & \multicolumn{2}{c|}{$\sqrt{s}=200$  GeV}
                     & \multicolumn{2}{c|}{$\sqrt{s}=500$  GeV}\\[2mm]
\cline{2-5}
\rule{0mm}{6mm} state
    & $\sigma$   & $\sigma_{\gamma}$ & $\sigma$ & $\sigma_{\gamma}$ \\[1.5mm]
\hline
\rule{0mm}{7mm}
$\bar{u} u \bar{s} s $                                      
    & 83.63(12) & 13.45(13) &  32.561(51) & 5.891(41) \\[1.5mm]
$\bar{u} u \bar{b} b $                                      
    & 86.565(95) & 14.39(11) & 38.475(45) & 7.004(36) \\[1.5mm]
$\bar{c} c \bar{d} d $                          
    & 84.41(12) & 13.742(76) & 33.150(92) & 5.969(41) \\[1.5mm]
$\bar{c} c \bar{b} b $                                      
    & 87.209(83) & 14.445(99) & 39.237(42) & 7.149(35) \\[1.5mm]
$\bar{t} t \bar{d} d $                                      
    & -- & -- & 0.903(1) & 0.0801(2) \\[1.5mm]
$\bar{t} t \bar{s} s $                                      
    & -- & -- & 0.903(1) & 0.0803(2) \\[1.5mm]
$\bar{d} d \bar{s} s $                                      
    & 79.658(95) & 9.77(11) & 29.812(39) & 4.159(34) \\[1.5mm]
$\bar{d} d \bar{b} b $                                      
    & 82.486(87) & 10.436(86) & 37.435(39) & 5.363(26) \\[1.5mm]
$\bar{u} u \bar{c} c $                                      
    & 87.06(13) & 15.98(12)  & 35.826(61) & 7.581(52) \\[1.5mm]
$\bar{u} u \bar{t} t $                                      
    & -- & -- & 0.9071(11) & 0.09440(21) \\[1.5mm]
$ \bar{u} u \bar{u} u $                                     
    & 42.465(45) & 7.794(35) & 17.410(27) & 3.684(19) \\[1.5mm]
$ \bar{c} c \bar{c} c $                                     
    & 43.323(44) & 7.935(27) &  18.076(23) & 3.867(14) \\[1.5mm]
$\bar{d} d \bar{d} d $                                      
    & 39.173(32) & 4.791(20) & 14.758(15) & 2.0706(76)\\[1.5mm]
$\bar{s} s \bar{s} s $                                      
    & 39.167(32) & 4.810(20) & 14.767(15) & 2.0769(76) \\[1.5mm]
$\bar{b} b \bar{b} b $                                      
    & 41.667(24) & 5.217(14) & 22.242(14) & 3.2142(67) \\[1.5mm]
\hline
\end{tabular}
\end{center}
\end{table}
}

The results for the neutral current channels of reactions (\ref{born})
and (\ref{brems}) are shown in Tables 5, 6 and 7. We list purely
leptonic channels in Table~5, semi-leptonic channels in Table~6 and
purely hadronic channels in Table~7. The cross sections in Tables~5, 6
and 7 are typically much smaller then those of Tables~3 and 4. Mass
effects on the other hand are bigger. The stronger dependence on
fermion masses can be explained as follows. The neutral-current
reactions are dominated by $s$-channel Feynman diagrams which contain
the propagator of a photon decaying into a fermion pair. This causes a
$\sim 1/s_{ff'}$ behavior of the matrix element squared and results in
a relatively high sensitivity to the fermion pair threshold
$s_{ff'}=(m_f+m_{f'})^2$.  There is a relatively big effect for
charged lepton pairs $\mu^+\mu^-$, $\tau^+\tau^-$ and much smaller
effect for quark pairs, except for $\bar{t}t$ of course. This is due
to the fact that there is no cut on the invariant mass of a charged lepton pair
in Eqs.~(\ref{cancuts}). Again we
observe an inverse mass effect in some channels, especially those
containing a neutrino pair.
The cross section of $\epm \ra \bar{d} d \bar{\nu}_e \nu_e$ is bigger than that
of $\epm \ra \bar{u} u \bar{\nu}_e \nu_e$ although the mass of the
$d$-quark is almost twice as big as that of the $u$-quark. The inverse
mass effect is caused by the invariant mass cut $m(q,q')> 10$ GeV
of~(\ref{cancuts}), which is more restrictive for lighter fermion
pairs than for heavier ones and by the fact that there is neither an
invariant mass cut nor an angular cut on a neutrino-pair. We observe
that the mass effects depend on cuts.  They may be quite different for
different choices of cuts.

Whether or not the mass effects will play a role in the analysis of
the future data depends mostly on the luminosity of a future linear
collider. If we assume a total integrated luminosity of $500 ~{\rm
fb}^{-1}$ they will certainly become relevant. Therefore it is better
to keep nonzero fermion masses in the calculation, or test the
massless fermion generators against a massive one for each given set
of kinematical cuts.

If we compare the channels containing a $\bar{b} b$-- or $\bar{c}
c$--pair with the channels containing lighter quark flavors, we see a
clear signal of the Higgs--strahlung reaction $\epm \ra ZH$, especially
at $\sqrt{s}=500$ GeV, which exceeds the $ZH$ production threshold for a
Higgs boson mass of 115 GeV.
In case of $\epm \ra \bar{b} b \bar{\nu}_e \nu_e,
\bar{c} c \bar{\nu}_e \nu_e $, we see also a signal of
a $W^+W^-$ fusion mechanism of the Higgs boson production. For the final
state containing a $\bar{b} b$--pair, the signal is visible already at
$\sqrt{s}=200$ GeV and it becomes much more pronounced at
$\sqrt{s}=500$ GeV.

As the threshold energy for $\epm \ra \bar{t} t \bar{t} t$ is bigger
than 500 GeV, we do not show its cross section in Table~7. However,
even at $\sqrt{s}=800$ GeV the cross section of $\epm \ra \bar{t} t
\bar{t} t$ without cuts is of the order of $10^{-3}$ fb.

\section{Conclusions}
We have developed a program package {\tt ee4f$\gamma$} which allows
us to calculate in an efficient manner any process $\epm \ra 4{\rm
f},\;4{\rm f}\gamma$ keeping nonzero fermion masses. Corresponding
results have been investigated and discussed for all channels at
reference energies $\sqrt{s}$ = 200 GeV and 500 GeV. For detailed
investigations of physics at a high luminosity linear collider like
TESLA, these mass effects should be taken into account. Nonzero
fermion masses also provide a physical regularization of matrix
elements which exhibit collinear singularities in the zero mass
limit. The results thus may be used as benchmarks for calculations
performed in the approximation of vanishing fermion masses which
allows one to perform calculations with much less numerical
efforts. Our results may be considered as part of the complete
$O(\alpha)$ corrections which we will need to understand physics at
future high energy linear colliders in a precise manner. The
calculation of the missing virtual corrections to $\epm \ra 4{\rm f}$
is one of the big challenges for the future.

\bigskip

{\bf Acknowledgment}

We thank Anja Werthenbach for carefully reading the manuscript.

\end{document}